\documentclass{iaus}
\usepackage{graphics}
  \checkfont{eurm10}
  \iffontfound
    \IfFileExists{upmath.sty}
      {\typeout{^^JFound AMS Euler Roman fonts on the system,
                   using the 'upmath' package.^^J}%
       \usepackage{upmath}}
      {\typeout{^^JFound AMS Euler Roman fonts on the system, but you
                   dont seem to have the}%
       \typeout{'upmath' package installed. iaus.cls can take advantage
                 of these fonts,^^Jif you use 'upmath' package.^^J}%
      }
  \else
  \fi

  \checkfont{msam10}
  \iffontfound
    \IfFileExists{amssymb.sty}
      {\typeout{^^JFound AMS Symbol fonts on the system, using the
                'amssymb' package.^^J}%
       \usepackage{amssymb}%
         \let\leq=\leqslant
         
      }{}
  \fi

  \IfFileExists{amsbsy.sty}
    {\typeout{^^JFound the 'amsbsy' package on the system, using it.^^J}%
     \usepackage{amsbsy}}
    {}

\newsavebox{\astrutbox}
\sbox{\astrutbox}{\rule[-5pt]{0pt}{20pt}}

\newcommand\etal{\mbox{\textit{et al.}}}

\title[Outskirts of Galaxy Clusters: intense life in the suburbs]
      {HIROCS $-$ a galaxy cluster survey at high redshifts}

\author[S. Falter {\it et al.\/}]%
{S. Falter$^1$\thanks{Email: falter@mpia-hd.mpg.de},
H.-J. R\"oser$^1$, H. Hippelein$^1$, C. Wolf$^2$ \and E. Bell$^1$}
\affiliation{$^1$Max Planck Institute for Astronomy, K\"onigstuhl 17,
D-69117 Heidelberg, Germany \\[\affilskip]
$^2$Dep. of Physics, Denys Wilkinson Bldg., Univ. of Oxford, Keble Road,
Oxford, OX1 3RH, UK}

\pubyear{2004}
\volume{195}
\pagerange{1--8}
\date{?? and in revised form ??}
\jname{Outskirts of Galaxy Clusters: intense life in the suburbs}
\editors{A. Diaferio, ed.}
\begin{document}

\maketitle

\begin{abstract}
Clusters of galaxies are the largest bound gravitational systems in the
universe. They outline the large scale structure and thus are able to
test predictions of cosmological models. Detailed studies of galaxy
populations in clusters at high z offer insights into galaxy evolution
as a function of look-back time. Unfortunately, large homogeneous
samples of galaxy clusters only exist at redshifts $<$ 0.5. There are
around 1000 candidate clusters with photometric or otherwise estimated
redshifts above 0.5, but spectroscopically confirmed clusters are still
scarce in this redshift regime. Some clusters with z\,$>$\,1 have
recently been identified from X-ray surveys.
\end{abstract}

\firstsection % if your document starts with a section,
\section{Survey outline}
\small
\begin{table*}
\begin{center}
\begin{tabular}{lcccc}
\hline
 Field name   & RA (J2000) & DEC (J2000) & Size & E(B$-$V) \\
\hline
 MUNICS-S2F1 & 03$\rm ^h$\,06$\rm ^m$\,41$\rm ^s$.0 &
 00$^{\circ}$\,01$^{\prime}$\,12$^{\prime\prime}$ & 3$^{\circ} \times$ 1$^{\circ}$
  & 0.080 mag \\
 COSMOS-10h  & 10$\rm ^h$\,00$\rm ^m$\,28$\rm ^s$.0 &
 02$^{\circ}$\,12$^{\prime}$\,21$^{\prime\prime}$ & 2$^{\circ} \times$ 1$^{\circ}$
  & 0.018 mag \\
% COMBO17-S11 & 11$\rm ^h$\,42$\rm ^m$\,58$\rm ^s$.0 &
% -01$^{\circ}$\,42$^{\prime}$\,50$^{\prime\prime}$ & 2$^{\circ} \times$ 1$^{\circ}$
%  & 0.016 mag \\
 CADIS-16h   & 16$\rm ^h$\,24$\rm ^m$\,32$\rm ^s$.0 &
 55$^{\circ}$\,44$^{\prime}$\,32$^{\prime\prime}$ & 3$^{\circ} \times$ 1$^{\circ}$
  & 0.006 mag \\
 HIROCS-22h  & 22$\rm ^h$\,00$\rm ^m$\,00$\rm ^s$.0 &
 02$^{\circ}$\,14$^{\prime}$\,20$^{\prime\prime}$ & 3$^{\circ} \times$ 1$^{\circ}$
  & 0.065 mag \\
\hline
\end{tabular}
\caption[]{\footnotesize HIROCS fields\,$-$\,coordinates, sizes and
galactic extinction from \cite{Schlegel98}.\small} \label{tab:tab1}
\end{center}
\end{table*}
\normalsize HIROCS ($=$ Heidelberg InfraRed/Optical Cluster Survey)
intends to homogeneously select candidate clusters with redshifts up to
1.5. Our survey is part of a large programme (MANOS $=$ MPIA for
Astronomy Near-infrared Optical Surveys) exploring the evolution of
galaxies. To reach out to redshifts of 1.5 we are observing in the
infrared (H) supplemented by four optical (B, R, i, z) bands. HIROCS
will cover 11\,$\square^{\circ}$ in four fields, three stripes with
3$^{\circ}$\,$\times\,1^{\circ}$ and one with
2$^{\circ}$\,$\times\,1^{\circ}$ (see Table~\ref{tab:tab1}). This
ensures a sufficient coverage of large scale structure features in the
foreseen redshift range. Furthermore, predictions for the number of
galaxy clusters above unity from \cite{Bart02} for the SDSS show that
our dedicated survey area should yield a statistically significant
cluster sample. They predict more than 5 clusters per $\square^{\circ}$
with z\,$>$\,1.

With one exception (CADIS-16h field) we have equatorial fields with few
bright stars enabling follow-up studies from both hemispheres with the
VLT and the LBT (available in the course of 2004). The right ascension
of the fields is distributed such that observations are possible during
the whole year. Sufficiently low and uniform galactic extinction over
the field area is another requirement.

The MUNICS and CADIS fields provide extensive information for fixing the
selection function of the reduced HIROCS filter set. In these fields we
have multi-color data sets as well as a large number of spectra for
QSOs, galaxies and stars. We chose the COSMOS-10h field because the
COSMOS survey will obtain high-resolution images from HST. The
HIROCS-22h field is a new field.

\small
\begin{table*}
\begin{center}
\begin{tabular}{cccccccc}
\hline
  & m$\rm^\star_{k-corr}$ & E-corr & m$\rm^\star$ & obsLim & $\Delta$t & S/N
  & $\Delta$t/$\square^{\circ}$ \\
\hline
 B$\rm_J$ & 29.0 & -5.975 & 23.1 & 24.1 & 1500 & 19.0 &  6 ksec \\
 R$\rm_J$ & 27.3 & -2.878 & 24.4 & 25.4 & 5000 &  5.3 & 20 ksec \\
 i        & 26.7 & -2.291 & 24.4 & 25.4 &      &      & $\sim$20 ksec \\
 z        & 25.0 & -2.480 & 22.5 & 23.5 & 6000 &  5.2 & 24 ksec \\
 H        & 21.2 & -1.348 & 19.9 & 20.9 & 3000 &  5.0 & 48 ksec \\
\hline
\end{tabular}
\caption[]{\footnotesize Limiting Magnitudes (col. obsLim) and exposure
times to cover 1\,$\square^{\circ}$ for the HIROCS filters. For the
calculation a standard E galaxy SED is scaled to produce an absolute
magnitude of an L$^\star$-galaxy in the Gunn r-band of
M$\rm_r$\,=\,$-$21.75\,mag (\cite[Garilli, Maccagni \& Andreon
1997]{Garilli97}, \cite[Paolillo, Andreon, Longo, \etal\
2001]{Paolillo01}). This spectrum is k-corrected (z\,=\,1.5) and
corrected for evolutionary effects after \cite{Poggianti97}. The
observations are dithered and split into short exposures. The summed
integration time per pointing is shown in the column $\Delta$t. The last
column incorporates the field of view of the detector and specifies the
required time to cover 1\,$\square^{\circ}$. We assume that one hour of
exposure time yields a net science exposure time of 3 ksec. Cosmology:
H$_{\circ}$\,$=$\,65, $\rm\Omega_M$\,$=$\,0.3,
$\rm\Omega_\Lambda$\,$=$\,0.7. \small}
\label{tab:tab2}
\end{center}
\end{table*}

\normalsize Our survey aims at a limiting magnitude of L$^{\star}\,+$\,1
for an elliptical galaxy at z\,$=$\,1.5 (5$\sigma$).
Table~\ref{tab:tab2} shows the limiting magnitudes and the required
exposure times.

The multi-colour-classification scheme developed for the CADIS and
COMBO-17 surveys (\cite[Wolf, Meisenheimer \& R\"oser 2001]{Wolf01})
will classify all objects and assign photometric redshifts. However, the
accuracy of the redshifts will be not as accurate though sufficient for
cluster detections.

\section{Cluster finding}
\normalsize We will employ a 3-dimensional cluster finding algorithm
based on finding overdensities in position on the sky and in redshift
space. According to its redshift z$_0$ derived from the
multi-colour-classification the projected Abell radius R$\rm _A$ for
each object is calculated and the number of objects N$\rm _{obj}$ within
R$\rm _A$ counted. Only objects with velocity differences below a given
threshold are added to N$\rm _{obj}$:
\begin{equation}
\rm \frac{v}{c} = \frac{\left( 1+z_1\right)^2-\left( 1+z_0\right)^2}
{\left( 1+z_1\right)^2+\left( 1+z_0\right)^2} \leq \left(
\frac{v}{c}\right)_{lim}
\end{equation}

Candidate galaxy clusters are found as excess numbers in N$\rm _{obj}$
which exceed values expected for a random object distribution. From
these excess numbers we define an overdensity parameter.

\begin{figure}
\vspace{5cm} \includegraphics{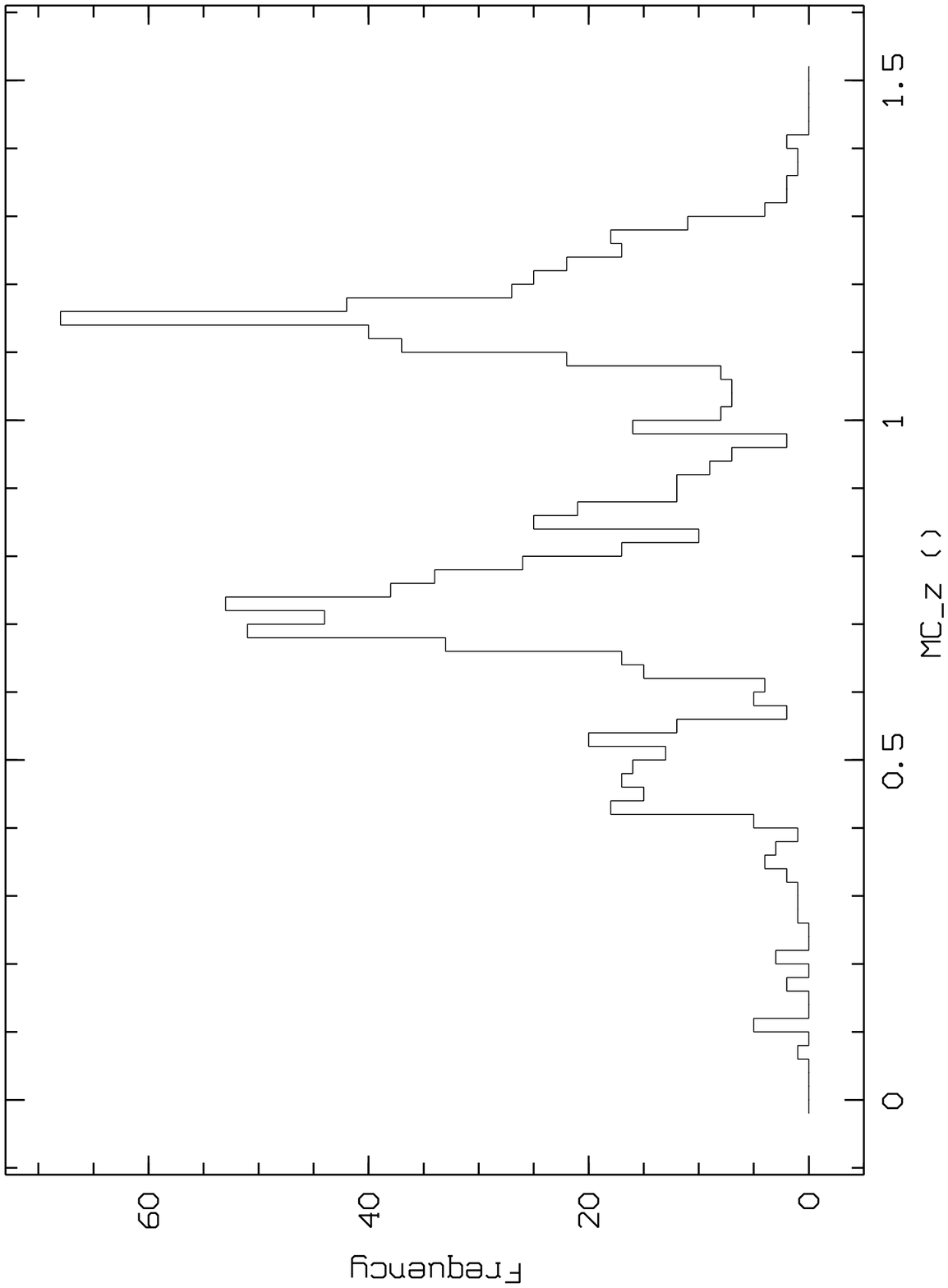} \includegraphics{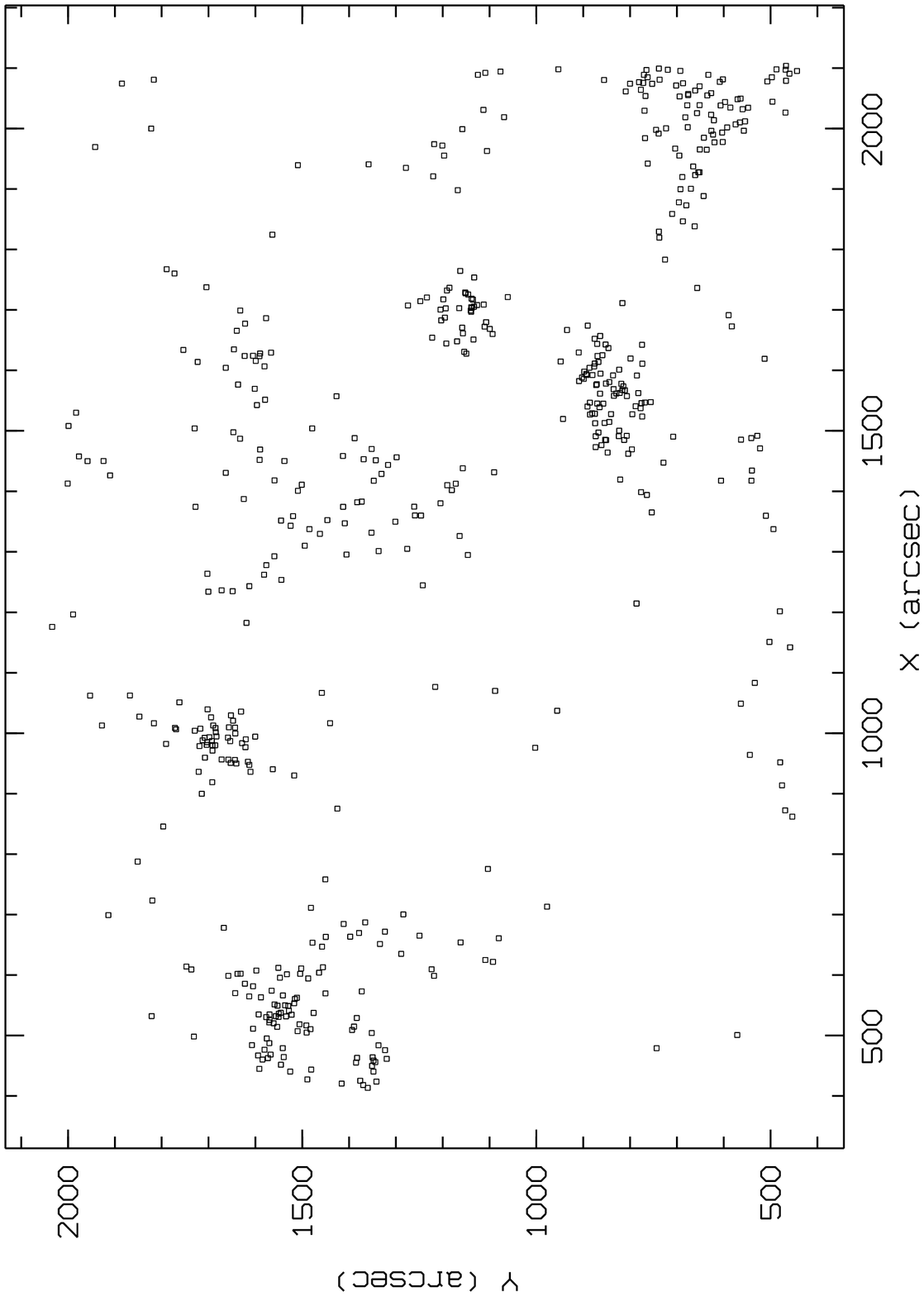} \caption{Left: z-histogram shows
object enhancements at z\,$\sim$\,0.5, 0.7, 1.1. Objects are selected by
overdensity; Right: Object distribution in the CDFS. The condensations
of points indicate candidate galaxy clusters (coordinates in arbitrary
units).} \label{fig:fig1}
\end{figure}

A first test application to the object catalog of the Chandra Deep-Field
South (CDFS) observed by COMBO-17 (field size
31$^{\prime}$\,$\times$\,30$^{\prime}$) reveals galaxy overdensities at
z\,$\sim$\,0.5, 0.7 and 1.1 (Figure~\ref{fig:fig1}).
%\begin{figure}
% \includegraphics{turin1.ps}
% \includegraphics{turin2.ps}
%  \caption{Left: z-histogram shows object enhancements at
%  z\,$\sim$\,0.5, 0.7, 1.1. Objects are selected by overdensity; Right:
%  Object distribution in the CDFS. The condensations of points indicate
%  candidate galaxy clusters (coordinates in arbitrary units).}\label{fig:contour}
%\end{figure}

In the past several other techniques for identifying clusters of
galaxies have been developed. We intend to compare the performance of
our algorithm with the Voronoi tessellation (\cite[Ramella, Boschin,
Fadda \& Nonino 2001]{Ramella01}), the matched-filter and the
friends-of-friends algorithm. Because of the imaging in five filters we
will also be able to apply the red sequence approach established for the
Toronto Red-sequence Cluster Survey (\cite[Gladders \& Yee
2000]{Gladders00}) for various colour combinations.

\section{Project status}

Since August 2003 we are collecting the survey data at the Calar Alto
3.5\,m telescope and since December 2002 with the Wide-Field-Imager
(WFI) at the ESO 2.2\,m telescope. A first impression of the quality of
the IR data gives a pipeline-reduced mosaic of 12 H-band images in the
HIROCS-22h field observed with the NIR wide-field camera OMEGA2000 on
this webpage:

http://www.caha.es/newsletter/news04a/OMEGA2000$\_$mosaic.pdf\\
It covers almost 3/4\,$\square^{\circ}$; the PSF is
$\sim$\,0.$^{\prime\prime}$8. Each frame is composed of 25 single
exposures with 1 minute exposure time, the limit is 19.9\,mag
(5$\sigma$). The holes around bright objects are due to over-corrected
background, as the pipeline, whose result is shown, uses only a local
sky determination.

\end{document}